\documentclass{rspublic}
\usepackage{graphics}
\usepackage{latexsym}
\usepackage{epsfig}

\newcommand{\ON}{{\sc on}}
\newcommand{\OFF}{{\sc off}}
\newcommand{\XOR}{{\sc xor}}

\newcommand{\tspr}{\tau_{\rm spr}}
\newcommand{\tloop}{\tau_{\rm loop}}
\newcommand{\ds}{\Delta_s}
\renewcommand\refname{\normalsize\textsf{References}}

\begin{document}

\title{On the Origin of Chaos in Autonomous Boolean Networks} 
\author[H. L. D. S. Cavalcante \textit{et al.}]{Hugo L. D. de S. Cavalcante, Daniel J. Gauthier,\\ Joshua E. S. Socolar, Rui Zhang}
\affiliation{Duke University, Department of Physics and Center for Nonlinear and Complex Systems, Durham, NC 27708 USA}

\maketitle

\begin{abstract}{Chaos, Boolean Networks, Time-Delay Dynamical Systems}
We undertake a systematic study of the dynamics of Boolean networks to determine the origin of chaos observed in recent experiments.  Networks with nodes consisting of ideal logic gates are known to display either steady states, periodic behavior, or an ultraviolet catastrophe where the number of logic-transition events circulating in the network per unit time grows as a power-law. In an experiment, non-ideal behavior of the logic gates prevents the ultraviolet catastrophe and may lead to deterministic chaos. We identify certain non-ideal features of real logic gates that enable chaos in experimental networks. We find that short-pulse rejection and the asymmetry between the logic states tends to engender periodic behavior, at least for the simplest networks. On the other hand, we find that a memory effect termed ``degradation'' can generate chaos. Our results strongly suggest that deterministic chaos can be expected in a large class of experimental Boolean-like networks. Such devices may find application in a variety of technologies requiring fast complex waveforms or flat power spectra, and can be used as a test-bed for fundamental studies of real-world Boolean-like networks.

\end{abstract}
\shorttitle{Boolean Chaos}

\section{Introduction}
Boolean models are often used to obtain insights into the dynamical properties of physical systems composed of elements that appear to execute binary logic.   A paradigmatic case is the behavior of digital circuits in which physical gates are designed with the specific intention of executing Boolean logic.  In typical circuits designed to carry out well defined computations, one typically introduces an external clock that determines when each gate is to be updated.  By making the time between ticks of the clock sufficiently long, it is possible to make devices that accurately carry out any desired Boolean operations (von Neumann 1956). 

Computation is not the only possible use for digital circuitry, however.  For some applications, such as private communications, remote sensing, or random number generation, one may want circuits that generate chaos with an ultra-broadband spectrum.  One approach to creating such circuits is to do away with the clock and allow each gate to respond continuously to its inputs.  The analog characteristics of the response at very high frequencies, along with the different signal propagation delays between the gates, can then lead to complicated dynamics that may or may not be captured by Boolean models.  We call such devices {\em autonomous digital circuits}.  We wish to understand the potential sources of chaotic dynamics in autonomous digital circuits and other physical systems involving interactions with similar characteristics.

An autonomous Boolean network (ABN) is a set of nodes with binary values coupled by links with associated time delays.  Each node is updated continuously according to a designated Boolean function of the values of its inputs at the appropriate previous times.  If node $A$ receives an input from node $B$, we refer to $A$ as a ``target'' of $B$.  In principle, the time delay between the switching of node $A$ and its target $B$ is due to a signal propagation time on the link, which may depend on whether the switch was a rise (from \OFF\ to \ON) or a fall (from \ON\ to \OFF).  It is also possible that the processing time at $B$ depends on the state of all the other inputs to $B$ at the time the signal arrives.  In general, we expect a nominal time delay associated with each link and slight adjustments depending on what information is being transmitted and whether it induces activation or decay of the target node.  Moreover, it is known that the set of attractors of a network can be influenced by memory effects (Norrell {\it et al.}\ 2007), which can be modeled in ABNs as a dependence of the time delays on the amount of time that the receiving node has been in its current state.

\begin{figure} 
 \begin{center}
\resizebox{\textwidth}{!}{\includegraphics{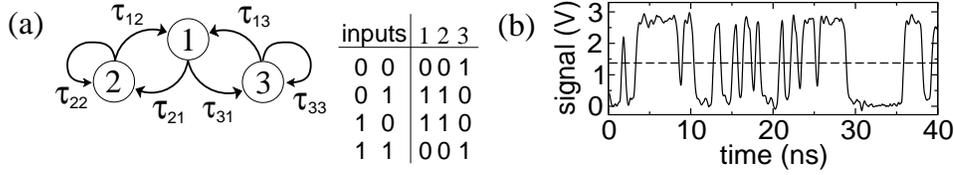}}
\end{center}
\caption{\label{fig:network} (a) Topology of the chaotic Boolean network investigated by Zhang {\it et al.}\ (2009) and truth table for logic operation performed by the nodes 1, 2 ({\sc xor}), and 3 ({\sc xnor}) on their respective inputs.
(b) Temporal evolution of the voltage at one point of the chaotic network.}
\end{figure}

The present communication is motivated by recent experiments by Zhang {\it et al.}\ (2009), who constructed an autonomous digital circuit using commercially-available, high-speed electronic logic gates.  The topology of their Boolean network is shown in Fig.\ \ref{fig:network}(a).  It consists of three nodes that each have two inputs and one output that propagates to two different nodes. The time it takes a signal to propagate to node $j$ from node $i$ is denoted by $\tau_{ji}$ ($i,j=1,2,3$).  Nodes 1 and 2 execute the Exclusive-{\sc or} ({\sc xor}) logic operation, while node 3 executes the {\sc xnor} (see truth tables in the Fig.~\ref{fig:network}(a)).  There is no clock in the system; the logic elements process input signals whenever they arrive, to the extent that they are able. They observed that the temporal evolution of the voltage at any given point in the circuit has a non-repeating pattern with clear Boolean-like state transitions, displays exponential sensitivity to initial conditions, and has a broad power spectrum extending from dc to beyond 2 GHz. Fig.~\ref{fig:network}(b) shows the voltage at the output of node 2. Because the circuit includes feedback loops with incommensurate time delays, it spontaneously evolves to dynamical states with the shortest possible pulse widths, a regime in which time-delay variations generate chaos. 

\begin{figure} 
\begin{center}
\resizebox{6cm}{!}{\includegraphics{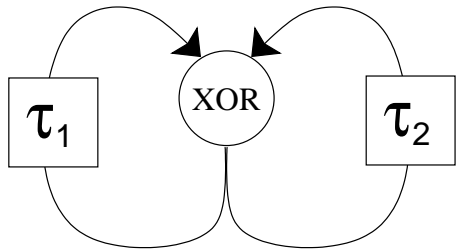}}
\end{center}
\caption{The network of primary interest in this paper.  
  \label{fig:xor}}
\end{figure}

Zhang {\it et al.}\ (2009) also demonstrated using numerical simulations that an ABN model can account for the major features observed in their network.  Our goal here is to use analysis and numerical simulations to identify the possible sources of chaos in simple ABNs and thereby clarify the origins of chaotic dynamics observed in autonomous digital circuits.  In this paper, we focus on the simplest network that yields nontrivial behavior -- a single \XOR\ gate with two output links that both feed back to its own inputs, shown schematically in Fig.~\ref{fig:xor}.

There are two generic effects in Boolean circuits that dramatically alter the attractor structure.  First, the time delay on the link from $B$ to $A$ depends on whether $A$ and $B$ are switching \ON\ or \OFF\ (Norrell {\it et al.}\ 2007), as mentioned above. 
We refer to the special case in which time delays do {\em not} depend on the direction of the switches as {\em symmetric}.  Second, the nodes in the network cannot process pulses of arbitrarily short duration; pulses shorter than some cutoff duration $\tspr$ are filtered out so that they never reach the next target node (Klemm \& Bornholdt 2005, Norrell {\it et al.}\ 2007). 
A pulse is defined as two consecutive state transitions of a single node. The pulse width is the temporal separation between those transitions. 
If a transition turning a node \OFF\ (or \ON) were to occur earlier than $\tspr$ after a transition turning that node \ON\ (or \OFF), the system evolves as if neither transition had ever occurred.  We refer to this effect as {\em short-pulse rejection} and assume it is present in all cases.  

We use the term {\em symmetric} ABN, or SABN, to refer to a system with a short-pulse rejection mechanism and time delays that are independent of input and target states and their histories.  In the limit $\tspr \rightarrow 0$, our SABN is equivalent to the Boolean Delay Equations discussed by Ghil and collaborators (Dee \& Ghil 1984, Ghil \& Mullhaupt 1985, and Ghil \textit{et al.} 2008).

A third effect that turns out to be quite important is the dependence of the time delay along a link on the state of the input node and its recent history.  This has been termed the ``degradation'' effect because it typically takes the form of a variation in delay time for switches at the trailing edge of pulses near the short-pulse-rejection limit (Bellido-D\'{i}az \textit{et al.} 2000).  We will see below that this effect is necessary and often sufficient to generate chaos in simple networks.

This paper is organized as follows.  In Section~\ref{sec:singleloop} we establish notation and some useful definitions and discuss the periodic behavior of networks with only a single feedback loop with no degradation effect. In Section~\ref{sec:twoloops}, we show rigorously that the symmetric two-loop \XOR\ system with no degradation can have only periodic (or fixed point) attractors.  In Section~\ref{sec:divergence}, we argue that chaotic behavior should not be expected in any system in the absence of a degradation effect, though we cannot rule out the possibility entirely.  In Section~\ref{sec:chaos}, we present a numerical model of a degradation effect in the simplest possible feedback system --- a single copier node with a self-input --- and show that it can produce chaos.  Finally, in Section~\ref{sec:degradation_xor}, we present numerical results elucidating the nature of the chaos that appears in the \XOR\ system when degradation effects are included.

\section{Single loops}\label{sec:singleloop}
Assuming that there are no time-varying external inputs to the network, a purely feed-forward network will have only fixed point attractors.  Any persistent periodic or chaotic oscillations in the system must be driven by some sort of feedback loop or combination of multiple feedback loops, where a feedback loop is defined as a ring of any number of nodes that allows a signal generated at one node to propagate back to the input of that node.  The simplest case is a single node that is its own target.

\begin{figure} 
\begin{center}
\resizebox{7cm}{!}{\includegraphics{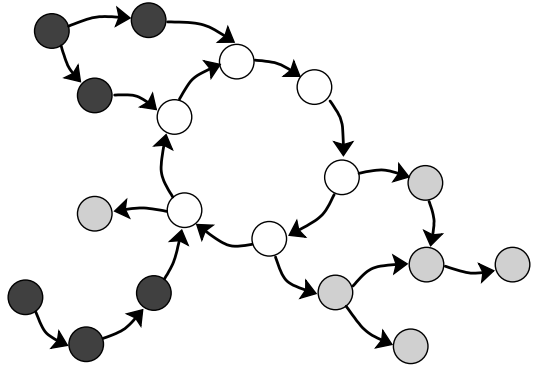}}
\end{center}
\caption{A network with one feedback loop.  White nodes form the loop.  Dark nodes may determine whether nodes in the loop act like copiers or inverters.  Light gray nodes are slaved to the dynamics of the loop and do not influence it in any way.
  \label{fig:singlering}}
\end{figure}

The basic structure of a network containing a single loop is one ring of nodes, each of which either copies or inverts its input.  The system may also contain additional nodes that are targets of one or more nodes on the ring, and these targets may themselves have additional targets forming feed-forward subnetworks that lead to dead ends.  Finally, the nodes on the ring may have additional inputs that are controlled by feed-forward chains.  These nodes may determine whether each node on the ring acts as a copier or an inverter.  (See Fig.~\ref{fig:singlering}.)

To explain the behavior SABNs containing only one feedback loop, we introduce some definitions and notation.  Consider a loop of $N$ nodes in which node $i$ is an input to node $i+1$ and node $N$ is an input to node $1$.  Let $\tau_{BA}$ designate the time delay associated with the link with input $B$ and target $A$.  We refer to each switch from \OFF\ to \ON\ in a time series of any given node as a {\em positive kink} and each switch from \ON\ to \OFF\ as a {\em negative kink}.  A kink may be thought of as propagating along a link and then being processed by the target node.  

A {\em copier} transmits the kink to its outputs and an {\em inverter} changes the sign of the kink before transmitting it.  At any given instant, there may be many kinks on the loop.  There is a topological constraint, however, depending on whether the loop has an even or odd number of inverters.  We use the terms  {\em even loop} and {\em odd loop} to distinguish these cases.   On an even loop, the single-valued nature of each node forces the number of kinks to be even.  On an odd loop, the number of kinks must be odd.  

To specify a state of the system, we must specify a continuous time series for each node over a time interval $[t-\tau_{i-1,i},t]$.  We will see later that this is important for measurements of trajectory divergence.  For now, we simply note that all $N$ such time series must be specified as an initial condition in order to determine the subsequent dynamics.  The system state specified by the initial conditions may be visualized as a number of kinks that are propagating along their respective links at time $t$ and will reach their targets before any signal from their input nodes can.
  
Let $\tloop\equiv \tau_{N,1}+\sum_{i=1}^{N-1}\tau_{i,i+1}$ be the sum of the time delays on all links in the loop.  For any initial condition, if $\tspr$ is set to zero, then every kink present moves around the loop in time $\tloop$.  On an even loop, the system will return to its original configuration at this time.  On an odd loop, the
configuration at $t=\tloop$ will be the inversion of the original and the system will be periodic with period $2\tloop$.  Setting $\tspr$ to a nonzero value results in the elimination of pairs of opposite sign kinks that are separated in time by less than $\tspr$.  Once all such pairs have been eliminated, short-pulse rejection mechanism plays no further role and the system is periodic.

Allowing different delay times on each link for kinks of different signs causes a dramatic reduction in the number of attractors.  The time required for a positive kink beginning at a given site to make a full circuit and return to that site will be different (in the absence of fine tuning) from the time required for a
negative kink to return.  Thus, if there are two kinks in the system, one will catch up to the other, eventually leading to a pulse of width smaller than $\tspr$, which will be annihilated.  In an even loop, all pulses will eventually annihilate and the attractors will always be fixed points in which all nodes hold steady values consistent with their input.  

The case of odd loops is more complicated.  When a positive kink propagates around the loop once, it is converted to a negative kink.  Thus, a kink of either sign will take exactly the same time to propagate around the loop twice.  If a pulse is wide enough to avoid annihilation during the time required for the two traversals, it will return with no change in its width.  A pulse of this type, however, is only marginally stable.  If the width of the pulse is perturbed, there is no mechanism for restoring it to its original value.  In a system where noise causes small random fluctuations in the delay times, the width of a pulse will execute a random walk and the pulse will eventually collapse due to short-pulse rejection.  Because of the topological constraint, there will always be one kink left that cannot be annihilated, and it will propagate, creating a unique oscillatory attractor. 

Thus, we see that SABNs are special in that they admit a large set of marginally stable attractors that collapse to a much smaller set for arbitrarily small symmetry breaking (in the even loop case) or noise (in the odd loop case).  We will return to the asymmetric case later.  For now, we continue the discussion of SABNs.

\section{Two loops}\label{sec:twoloops}
We next consider the simplest possible (nontrivial) system with two feedback loops: a single \XOR\ gate whose two inputs come directly from its own output, shown earlier in Fig.~\ref{fig:xor}. The output of an \XOR\ gate changes its value every time one of its inputs changes, so that kinks propagating through this network can be annihilated only through short-pulse rejection. If the \XOR\ is replaced with any two-input logic function other than {\sc not xor}, the dynamics leads quickly to a fixed point or a very simple oscillation, as can be checked by inspection.  The {\sc not xor} case is identical to the \XOR\ under exchange of the meaning of \ON\ and \OFF. 

The dynamics of the system is defined by a Boolean delay equation for the state $x(t)$, together with a procedure for rejecting short-pulses. 
 The Boolean delay equation is  
 \begin{equation}
 x(t) = x(t-\tau_1) \oplus x(t-\tau_2).
 \label{eq:xor}
 \end{equation}
Let $\epsilon$ be an infinitesimal duration.
To implement short-pulse rejection, we adjust $x(t)$ as follows.  If $x(t) \neq x(t-\epsilon)$, indicating that a switch has occurred at time $t$, and $\int_0^{\tspr} x(t-s) \oplus x(t-\epsilon) ds \neq 0$,
indicating that the gate has switched sometime during the past short-pulse rejection interval, then $x(t-s) = x(t-\tspr)$ for all $s$ in $[0,\tspr)$. 

In the two-loop case, it is not obvious that the attractors must be periodic.  Ghil and co-workers have studied the case of no short-pulse rejection and noted that the system exhibits an ultraviolet catastrophe in which kinks become dense in time and pulse widths tend toward zero (Ghil \& Mullhaupt 1985, Ghil \textit{et al.} 2008).  The short-pulse rejection mechanism in our SABNs eliminates kinks that are too close and thereby regularizes the divergence.  The following theorem shows that what remains can only be periodic. We count the trivial always-\OFF\ fixed point as a degenerate case of a periodic attractor.

\begin{theorem}\label{thm:xor} 
For a SABN consisting of a single \XOR\ with two self-inputs having delays $\tau_1$ and $\tau_2$ (as shown in Figure~\ref{fig:xor}), the attractors are always periodic.
\end{theorem}

{\em Proof:} We will first show that the attractor reached when the system is initiated with a single kink is always periodic.  We will then show that the introduction of additional kinks in the initial condition cannot alter this result.

Assume, without loss of generality, that $\tau_1<\tau_2$.  Let $t_s$, for $s=1,2,\ldots$ represent the times that the output of the gate switches, and define the intervals between switching times as $\ds\equiv t_s-t_{s-1}$.  Now note that the future of the system is determined if a past sequence of switching events spanning a duration of $\tau_2$ (the extent of the memory encoded in the longest delay line) is specified.  The strategy is to show that the set of possible sequences $\ds$ for $s_1<s<s_2$ is finite for values of $s_2-s_1$ corresponding to $t_{s_2}-t_{s_1}<\tau_2$.  If this is true, the system must eventually revisit some sequence that is long enough to determine its future behavior, which immediately implies periodicity.  (Any deterministic system with a finite number of states must have only periodic or fixed point attractors.)

\begin{figure} 
\begin{center}
\includegraphics[scale=0.75]{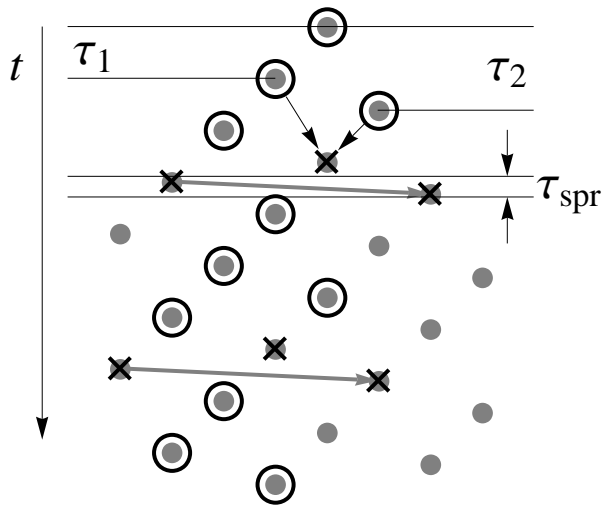}
\end{center}
\caption{Example of a lattice of possible switching times and the
  pattern of actual switching times after short-pulse rejection
  effects are taken into account. The vertical spacings corresponding
  to $\tau_1$, $\tau_2$, and $\tspr$ are shown.  The two arrows near
  the top of the lattice indicate two kinks that arrive simultaneously
  at a dot and therefore produce no outgoing kink.  The grey vectors
  indicate minimal rejection pairs.  The circled dots represent the
  events that actually occur.\label{fig:lattice}}
\end{figure}

Consider a system initialized by a single kink at time $t_0=0$.  That is, the gate is assumed to be \OFF\ for all times less than $t_0$ and switched on at $t_0$.  Each time the kink propagates around one of the delay lines, it causes the output of a new kink.  Thus, kinks could conceivably be generated at times 
\begin{equation}\label{eqtn}
t_{i,j} = i\tau_1 + j\tau_2
\end{equation} 
for any positive integers $i$ and $j$.  It is convenient to represent these times as sites of a lattice -- the dots in Fig.~\ref{fig:lattice} (Ghil \& Mullhaupt 1985). 
 The vertical axis in the figure represents time.  The horizontal axis is not physical.  It simply provides a way to visualize the causal processes that generate kinks.  Moving down and to the left from a given site leads to another site a time $\tau_1$ later.  Moving down and to the right leads to another site a time $\tau_2$ later. Note that the full set of sites is an infinite wedge subset of a Bravais lattice.  We refer to a given site and its corresponding event by the pair $(i,j)$ that specifies the time the event occurs according to Eq.~(\ref{eqtn}).

Event $(i,j)$ may {\em not} actually occur because some kinks are annihilated by the short-pulse rejection mechanism.  If events are generated at $(i,j)$ and $(k,\ell)$ for which $|t_{i,j}-t_{k,\ell}|<\tspr$, a pulse will be created that is too short to pass through the gate.  Thus, those two events will not generate any future events.  To trace out the dynamics on this lattice, we begin by circling the top site, $(0,0)$.  We then circle $(0,1)$ and $(1,0)$, indicating that events will occur at the corresponding times.  For each event $(i,j)$ that occurs, we circle the two events $(i,j+1)$ and $(i+1,j)$.  The sites are circled in chronological order, and, if the time interval between a newly circled site and the last one circled is less than $\tspr$, both circles are removed, indicating that neither event actually occurs.  A degenerate case arises when a single site gets circled twice -- once from each of its upstream neighbors.  This represents two kinks arriving simultaneously, which does not cause a switch in the output of the \XOR\ gate, so the site does not get circled.  Figure~\ref{fig:lattice} shows an example.

We now show that there is an upper bound to the horizontal distance
between two circled sites that are vertically separated by less than
$\tau_2$.  That is, if $|t_{i,j}-t_{k,\ell}|<\tau_2$, then
$|k-i|+|j-\ell|$ is bounded from above.  Let $D$ be the smallest
horizontal distance between two events that are vertically separated
by less than $\tspr$; \textit{i.e.}, $D$ is the smallest value of $m+n$ for
which $|m\tau_1-n\tau_2|<\tspr$.  In the case shown in
Fig.~\ref{fig:lattice}, $D=5$ (from $m=3$ and $n=2$).  The vector
joining two such sites is $\Vec{v}_{\rm min}\equiv(D,m\tau_1-n\tau_2)$.
We will refer to a pair of sites separated by $\Vec{v}_{\rm min}$ as a
``minimal rejection pair.'' 

If $\tau_2/\tau_1$ is rational, the lattice contains multiple sites that occur at exactly the same time and are separated by a minimal horizontal distance $W$.  ($W$ is equal to $D$ if and only if $\tspr$ is sufficiently small.)  We may identify such points, thus turning the lattice into a strip of width $W$ having periodic boundary conditions -- a cylinder of circumference $W$.  At any given time $t$, the configuration of circled sites lying in a band between $t-\tau_2$ and $t$ determines the future evolution uniquely.  Without loss of generality, assume that $t$ coincides with a lattice site.  Because the sites form a Bravais lattice, the configuration of lattice sites within the band is finite (and unique).  The number of possible configurations of circled sites is therefore finite, which ensures that the system must eventually revisit some configuration that it has already passed through.  The subsequent evolution will then cycle periodically through that configuration.

For the case of irrational $\tau_2/\tau_1$, there are no pairs of sites that occur at exactly the same time.  Nevertheless, we can prove that the set of circled sites must be confined to a region whose width never exceeds $D$, so the number of accessible configurations is again finite.  The proof proceeds by contradiction.  Assume that there are two circled sites separated vertically by less than $\tau_2$ and horizontally by more than $D$.  Because no two sites sit at exactly the same time, each circled site must be connected to $(0,0)$ by an unbroken chain of circled sites that caused it.  But any two such chains that begin from sites separated by more than $D$ must contain a minimal rejection pair.  To see this, consider any two paths starting from the same site.  
Let $(i_r,j_r)$ with $r=0,1\ldots R$ denote the sites on one path, and $(k_s,\ell_s)$ with $s=0,1\ldots S$ be sites on the other.  Define $(\mu,\nu)_{r,s}\equiv(k_s-i_r,\ell_s-j_r)$, where $(\mu,\nu)_{0,0}=(0,0)$. Each step on either trajectory either changes $\mu$ or $\nu$ by $\pm 1$.
Thus, the set $\{(\mu,\nu)\}$ must contain every possible pair with $\mu < k_s-i_r$ and $\nu<\ell_s-j_r$. Now, assume that the paths contain sites corresponding to times that differ by less than $\tau_2$ and having a horizontal separation greater than $D$.  Let $(i_R,j_R)$ be the site on the left and $(k_S,\ell_S)$ be the site on the right, so that $i_R-k_S$ and $\ell_S-j_R$ are both positive.
We then have
\begin{equation} \label{eqn:horizontalsep}
(i_R-k_S) + (\ell_S-j_R)>D
\end{equation}  
and
\begin{equation} \label{eqn:verticalsep}
|(i_R-k_S)\tau_1-(\ell_S-j_R)\tau_2| < \tau_2.  
\end{equation}  
Recall that $(m,n)$ gives the minimal rejection pair.  If $(i_R-k_S)<m$, Eq.~(\ref{eqn:horizontalsep}) requires $(\ell_S-j_R)>n$, which implies that Eq.~(\ref{eqn:verticalsep}) must be violated, as can be seen immediately by comparing to the known relation $m\tau_1-n\tau_2<\tspr$, with $\tspr<\tau_2$.  Similarly, $(\ell_S-j_R)<n$ requires $(i_R-k_S)>m$, which again implies a violation.  Thus, we must have $(i_R-k_S)\geq m$ and $(\ell_S-j_R)\geq n$, which implies in turn that there must be some pair $(r,s)$ for which $(\mu,\nu)_{r,s} = (m,n)$.  When these sites were circled, however, they would have been subject to short-pulse rejection.  Hence the configuration of two chains cannot represent a possible trajectory of the system.

We have proven that the dynamics initiated by a single kink must be confined to a tube of width $D$ on the lattice.  The tube need not be vertical or even straight, but it cannot have a horizontal width greater than $D$ at any time.  The periodicity of the trajectory then follows immediately from the fact that the number of configurations of circled dots that can be covered by a rectangle of height $\tau_2$ and width $D$ is finite, which guarantees that some configuration will be repeated after a sufficiently long time, and the trajectory between these repeated configurations will then be repeated {\it ad infinitum}.  Figure~\ref{fig:lattice} shows an example in which the trajectory repeatedly returns to a configuration in which only one kink is present.

To complete the proof that all attractors are periodic, we must consider the possibility of initiating the system with two or more kinks.  To analyze such cases, we overlay the lattices originating from each kink in the initial interval of duration $\tau_2$.  The $(0,0)$ sites from the different lattices will be displaced vertically by times between $0$ and $\tau_2$.  The lattices emanating from each $(0,0)$ site will all be translated copies of the same lattice.

To see that the entire set of events must still be confined to a tube of finite width, first note that any event must be connected to one of the $(0,0)$ sites by a chain of events contained entirely in one lattice.  By the reasoning used in proving Theorem~\ref{thm:xor}, the events on any single lattice must be confined to a tube of width $D$.  The problem, then, is to show that there is an upper bound on the horizontal distance between circled sites on two different lattices when their vertical separation is less than $\tau_2$.

If $\tau_2/\tau_1$ is rational, the lattices can all be represented on a finite radius cylinder (or infinite plane with periodic boundary conditions at the sides) and it is clear that the number of configurations in a band of any given height is finite.  If $\tau_2/\tau_1$ is irrational, the proof follows essentially the same reasoning as the single lattice case.  Consider events on any two of the lattices.  There are now two minimal rejection vectors $\Vec{v}_{\rm min}$ depending on which lattice contains the trajectory on the right (with larger $n-m$).  Nevertheless, given any two widely separated points, the paths to each point that must pass through one of the minimal rejection separations for that pair of lattices.  This must be true for each distinct pair of lattices, so the entire trajectory must be confined to a finite tube.  Because short-pulse rejection prevents initialization with arbitrarily close kinks, the number of initial kinks is bounded.  This means that the number of overlayed lattices is bounded, which implies again a finite number of configurations within a rectangle of height $\tau_2$ and a specified width, so the full trajectory must be periodic. {\bf Q.E.D.}

\begin{theorem} 
For a SABN consisting of a single \XOR\ with two self-inputs having delays $\tau_1$ and $\tau_2$, and with $\tspr$ sufficiently small that no collapse to the always-\OFF\ state occurs before $t = \tau_2$, the trajectory will never reach the always-\OFF\ state. \label{thm:nocollapse}
\end{theorem}

\begin{figure} 
\begin{center}
\resizebox{10cm}{!}{\includegraphics{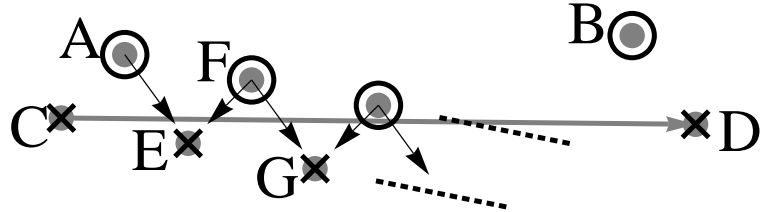}}
\end{center}
\caption{
  A portion of the event lattice relevant for the proof of
  Theorem~\ref{thm:nocollapse}.  The long grey vector indicates
  short-pulse rejection.  The other arrows show events required to
  annihilate additional kinks generated from events $A$ and $B$.
\label{fig:nocollapse}}
\end{figure}
{\em Proof:} The only way to reach the always-\OFF\ state would be for the two last events to annihilate by short-pulse rejection. That is, for the pattern of circled lattice sites to produce two candidates at the minimal rejection distance without producing any other circled dots on the interior of the minimal width tube
containing the trajectory.  Figure~\ref{fig:nocollapse} shows that this cannot happen.  When event $A$ generates event $C$, it must also generate $E$, which occurs later than $C$ and $D$. (We assume that $E$ and $D$ are distinct points. If $\tspr$ is too large, the collapse can occur immediately because $C$ and $D$ are both derived from $A$). In order for $E$ to be annihilated, event $F$ must be present, which in turn would generate $G$.  Note that $F$ has the same value of $m+n$ as $A$.  Repeated application of this reasoning shows that annihilation of all further events through exact coincidences would require an infinite line of events at the same value of $i+j$ as $A$, but such a line is impossible, both because it would eventually reach the edge of the triangular wedge of possible lattice points and because it would require events occurring at later times than our supposed final annihilation of $C$ and $D$.  Thus, the always-\OFF\ collapse cannot occur. {\bf Q.E.D.}

We now turn to the asymmetric case, in which the time delays associated with positive ($\tau_1^{\mathrm{ON}}$ and $\tau_2^\mathrm{ON}$) and negative kinks ($\tau_1^{\mathrm{OFF}}$ and $\tau_2^{\mathrm{OFF}}$) on any given link may be different.  
The evolution is calculated as follows. We construct a queue of times at which the gate switches states. Let $t_1$ be a time at which the gate switches from $x$ to $X$. When $t_1$ is the earliest time in the queue, it is removed (processed) and two future times $t_1+\tau_1^{X}$ and $t_1+\tau_2^{X}$ are added.  Let the next time after $t_1$ in the queue be $t_2$.  The switch at $t_2$ causes the gate to return to the state $x$. If $t_2-t<\tspr$, then $t_2$ is removed from the queue along with any times just added due to the switch at $t_1$.  Otherwise, only $t_2$ is removed and $t_2+\tau_1^{x}$ and $t_2+\tau_2^{x}$ are added. If $t_2+\tau_i^{x} < t_1+\tau_i^{X}$, however, then both are removed from the queue, as this implies that the trailing edge of a pulse overtook the leading edge as it propagated along the link.

The lattice picture now becomes more complicated.  We need a 4D lattice with one basis vector for each possible delay time.  Each time a site is circled, the state of the \XOR\ gate determines which two lattice directions are available for the next step, and the state of the gate is determined by the parity of the number of steps that have been taken up to the time in question.  Note that the number of steps cannot simply be counted by tracing the single path leading to the transition of interest.  All of the events above the one under consideration must be counted to determine the current state.

\begin{figure} 
\resizebox{\textwidth}{!}{\includegraphics{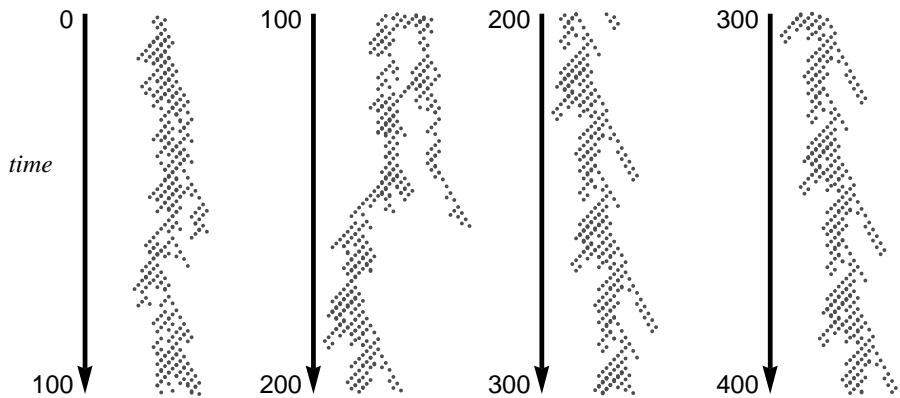}}
\caption{Typical trajectory of a two-loop ABN initialized with a single positive kink.
The columns are successive time intervals of 100 units and the periodic
attractor is evident in the fourth column.  The time delays for this simulation are $\tau_1^{\mathrm{ON}}=1.000$,  $\tau_1^{\mathrm{OFF}}=1.117$, $\tau_2^{\mathrm{ON}}= \phi$, and $\tau_2^{\mathrm{OFF}}=1.117 \phi$, where $\phi$ is the golden ratio, $1.618\ldots$.  The short-pulse rejection time was $\tspr = 0.040$. The system was initialized with a single positive kink at one of the inputs at $t=0$.
\label{fig:abn}}
\end{figure}

The proof of the bounded width of a trajectory on the 2D lattice breaks down for higher dimensions.  The difficulty is that it becomes possible for trajectories to avoid hitting pairs of points at the minimal rejection distance by moving in the third (or fourth) dimension.  We do not (yet) have a proof that periodicity is necessary for the asymmetric case, but extensive numerical simulations have failed to turn up any counterexamples.  Figure~\ref{fig:abn} shows a typical trajectory with the 4D lattice projected onto a plane.  A step one unit to the right indicates traversal of link 1, with the two possible vertical displacements $\tau_1^{\mathrm{ON}}$ or $\tau_1^{\mathrm{OFF}}$, similarly, steps to the left correspond to traversing link 2, with delay either $\tau_2^{\mathrm{ON}}$ or $\tau_2^{\mathrm{OFF}}$.  Only the circled sites are shown. 

We also could not prove that the asymmetric system will never collapse
to the always-\OFF\ state.  We cannot rule out the possibility that
two short-pulse rejections can annihilate all four of the events
emanating from two events separated by more than the minimal
short-pulse rejection distances on the 4D lattice.

\section{Trajectory divergence}\label{sec:divergence}
In the absence of a proof that all attractors of ABNs are periodic, it is useful to ask how one might detect chaos in a simulation.  Spectral analysis of the time series may be useful, but it is also possible for the periodic attractors to be extremely long and it may be difficult to resolve the peaks.  We consider two methods for following the divergence of trajectories that initially differ by a small perturbation.  In both cases, we introduce a perturbation by artificially delaying (or accelerating) the arrival of one kink by a small amount $\epsilon$.  In the first method, we construct the sequence of switching times $t_n$ for the original trajectory and $t'_n$ for the perturbed one.  We then plot $\delta_n=|t'_n-t_n|$ vs.\ $t_n$.  In the second, we define the Boolean difference at a given time to be 0 if the gate is in the same state on both trajectories and 1 if the states are different.  We integrate the Boolean difference between the original and perturbed trajectories over a time window $\tau_2$ (the longest delay time on a link). We denote the integral by $d(s)$, where $s$ represents the time between the applied perturbation and the end of the integration interval. 

It is immediately clear that the only source of trajectory divergence as measured by the first method is the rejection of a short-pulse in one trajectory but not the other.  There is no other mechanism that increases or decreases the time difference between corresponding kinks.  If there is a short-pulse rejection in one system but not the other, however, the pulse that makes it through could generate subsequent events that cause the two trajectories to diverge.  In the limit of infinitesimal $\epsilon$, short-pulse rejection differences will never occur and there can be no trajectory divergence.

In the second method, the distance between trajectories can grow because several kinks emanating from the perturbed one will also be perturbed by $\epsilon$.  This distance cannot grow larger than $n \epsilon$, where $n$ is the number of kinks that can be accommodated in an interval of duration $\tau_2$.  Note that $n$ is finite for any nonzero value of $\tspr$, so again there is no exponential divergence in the limit of infinitesimal $\epsilon$.  Further divergence requires short-pulse rejection differences as in the first method.

On the other hand, for any given $\epsilon>0$, there may be a pulse eventually generated with duration closer than $\epsilon$ to the short-pulse rejection threshold.  This effect could conceivably lead to exponential divergence in $d$ over an intermediate scale between the short-pulse rejection time and the time required for a kink to generate $n$ new ones --- roughly $n \tau_1$.  

The growth rate of the number of kinks generated by a single kink or pulse injected into a SABN is known to be polynomial when there is no short-pulse rejection, with an exponent of $\ell-1$, where $\ell$ is the minimum number of incommensurate delay times necessary to express the network evolution (Ghil \& Mullhaupt 1985). 
Short-pulse rejection will reduce the rate of growth as the number of kinks increases, however, so there will be no exponential divergence in SABNs.

\begin{figure}[htb] 
\begin{center}
\resizebox{8cm}{!}{\includegraphics{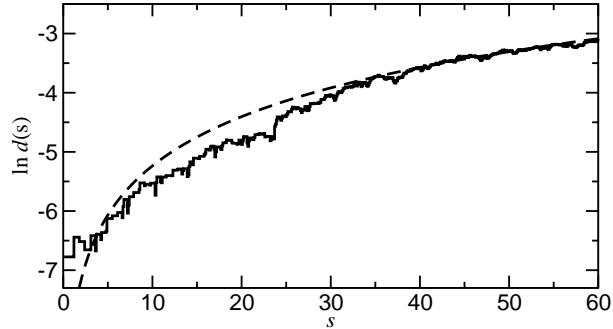}}
\end{center}
\caption{An average of $\ln d(s)$ for the \XOR\ gate with two delayed self-inputs and short-pulse rejection. The solid line is simulation data and the dashed line a fit to a power-law form of $d(s)$ with exponent 1.2.
\label{fig:polynomial_growth}}
\end{figure}

In the asymmetric case, we study numerically whether a single short-pulse rejection occurring in one trajectory but not the other can lead to exponential divergence in $d$ over an intermediate scale.  
Figure~\ref{fig:polynomial_growth} shows measurements of $\ln d(s)$ obtained from simulations with $\tspr = 1.0\times10^{-3}$, where a perturbation $\epsilon = 0.5\times 10^{-3}$ is applied at $t_0$.  
The transition time $t_0$ is chosen so that the original pulse covering $t_0$ was just filtered by the short-pulse rejection, while the pulse with transition time $t_0+\epsilon$ just barely survived. We average $\ln d(s)$ over 30 pairs of trajectories with different initial conditions. As we see in Fig. \ref{fig:polynomial_growth}, the distance grows polynomially in this case also. 
In section \ref{sec:twoloops}, we have proven that the SABN consisting of a single \XOR\ gate with two self inputs must yield only periodic attractors, and we have numerical evidence suggesting that, in the absence of memory effects (other than short-pulse rejection), the ABN will not yield exponential divergence of trajectories either.

\section{Boolean chaos}\label{sec:chaos}
The next place to look for a source of chaos in the autonomous digital circuit is in the memory effects associated with the response of the gate to two successive kinks that form a pulse. In principle (and in the circuit of Zhang {\it et al.}\ 2009), the delay time for a kink that is the trailing edge of a pulse can depend on the width of the pulse.  This dependence is called the {\em degradation effect} (Bellido-D\'{i}az \textit{et al.} 2000). 

We now present numerical evidence showing that the degradation effect can produce chaos in the simplest of all networks, a single copier with one self-input.  The dependence of delay times on pulse width introduces nontrivial dynamics in the durations of successive pulses which can lead to stabilization of pulse widths and numbers of kinks, and opens the possibility of chaotic sequences of pulse widths.

\begin{figure}
\resizebox{\textwidth}{!}{\includegraphics{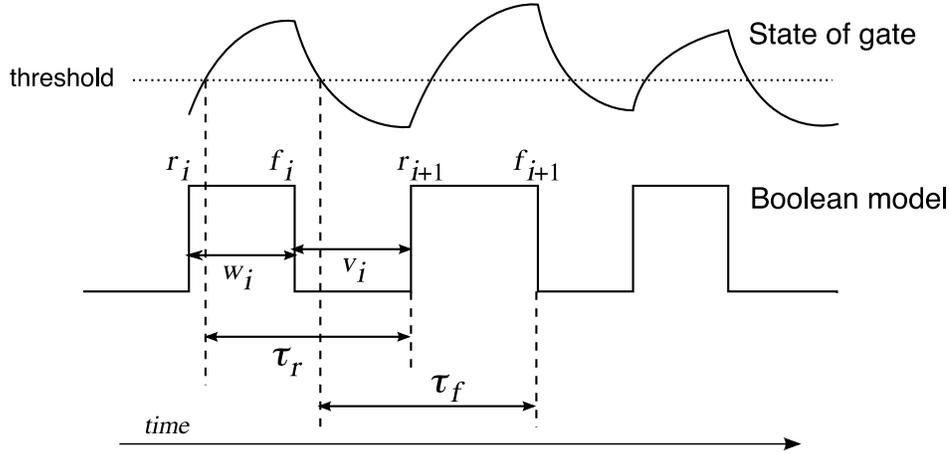}}\\ \\ 
\caption{Implementation of the degradation effect.  The top curve is a schematic
illustration of the temporal evolution of a state variable.  When the variable is above a threshold, the gate is considered to be \ON.  The lower curve
shows the Boolean approximation of the input to the gate.  For simplicity of illustration, we assume the gate is its own target.  The rising kink $r_i$ causes the state variable to begin a transition.  The indication that it has crossed the threshold is transmitted through a link with time delay $\tau_r$ and generates the rising kink at the input at $r_{i+1}$. The situation for falling kinks is similar.
\label{fig:degradation}}
\end{figure}

To understand the origin of the degradation effect requires that we consider
the underlying analog signal that has finite rise and fall times between \ON\
and \OFF\ states; the inherent propagation delays of signals propagating
through the logic gates; and the associated Boolean idealization of the
waveform. See Fig.~\ref{fig:degradation}.  

One method for implementing the degradation effect in the autonomous
Boolean network is as follows. Let $r$ be the time of occurrence of a rising kink and $f$ be the time of the subsequent falling kink.  Let $r'$ and $f'$ be the times of the rising and falling kink induced by the kinks at $r$ and $f$.  These times represent the arrival of a kink at the gate.  The actual time when the gate variable switches is slightly later and depends on the state of the gate at the time of the arrival of the kink.  That is, the actual switching time associated with the kink at $r$ is $r'-\tau_r$, where $\tau_r$ is the propagation delay. The difference between between $f'-\tau_f$ and $f$ is a function of the time that the gate was \ON; \textit{i.e.}, a function of $f-(r'-\tau_r)$.  We define a degradation function $g\left(f-(r'-\tau_r)\right)$ that gives the delay $f'-f$.  

For the case of a single pulse cycling through a single copier with a
self-input, we consider the evolution equations for the sequences $r_i$
and $f_i$.  
If the kinks are not rejected because of a short-pulse, $f_i$ will
generate a new kink $f_{i+1}$, with  
\begin{equation} \label{eqn:fj}
f_{i+1} = f_i + g_f(f_i - (r_{i+1}-\tau_r)).
\end{equation} 
Similarly, for a rising kink, we have
\begin{equation} \label{eqn:rj}
r_{i+1} = r_i + g_r(r_i-(f_i-\tau_f)).
\end{equation}
In principle, the constants $\tau_r$ and $\tau_f$ need not be equal
and the functions $g_f$ and $g_r$ may be different.  Because we are
interested only in showing the existence of chaos, we will consider only
the most tractable case: $\tau_r = \tau_f \equiv \tau$ and $g_r = g_f \equiv g$.

It is convenient to rewrite the evolution in terms of the pulse widths.
Defining $w_i\equiv f_i-r_i$ and $v_i\equiv r_{i+1}-f_i$, Eqs.~(\ref{eqn:fj}) and~(\ref{eqn:rj}) give
\begin{eqnarray}
w_{i+1} & = & w_i + g(\tau-v_i) - g(\tau-w_i)\,, \label{eqn:wv}\\
v_{i+1} & = & v_i + g(\tau-w_{i+1}) - g(\tau-v_i)\,. \label{eqn:wv2}
\end{eqnarray}
From Fig.~\ref{fig:degradation}, one sees that $w_i$ determines $v_i$ in a manner required by 
Eq.~(\ref{eqn:rj}): $g(\tau-w_i) - w_i = v_i$, which implies a constraint on the possible initial values and allows the map to be reduced to a single variable, say $v$.
Substituting this relation into the expression for $w_{i+1}$ (Eq.~(\ref{eqn:wv})) and inserting the results into Eq.~(\ref{eqn:wv2}) gives
\begin{equation} \label{eqn:vmap}\\
v_{i+1} =  v_i + g(\tau + v_i - g(\tau-v_i)) - g(\tau-v_i) \equiv h(v_i)\,.
\end{equation}
A stable fixed point of the map (\ref{eqn:vmap}) corresponds to periodic behavior in
which every pulse has the same width $w^*$ and every dip has the same
width $v^*$.  Equation~(\ref{eqn:wv}) implies $g(\tau-v^*) = g(\tau-w^*)$,
but this does not necessarily imply $w^* = v^*$.

The function $g(x)$ should satisfy two constraints.  First, $g(x)\geq\tau$ for all $x$
because $\tau$ is the minimum delay time required for traversing the link.
Second, $g(x)$ should asymptote to some constant $\tau_0>\tau$ for large $x$
because the gate will settle into a fixed state if its input is held constant for 
a long enough time.  By trial and error (based on intuition gleaned from experimental
studies of the response of real electronic gates), we
identify a function $g(x)$ that generates chaos in the simple copier model considered here.
It is given by
\begin{equation}
g(\tau-v) = \tau +a + b\left(\tau-v-c\right)\exp\left(-(\tau-v)/A\right),
\label{eq:g_of_x}
\end{equation}
with $\tau = 1.3$, $a=0.26$, $b=13.0$, $c=0.02$, and $A=0.18$. 
\begin{figure}[htb]
\begin{center}
\resizebox{10cm}{!}{\includegraphics{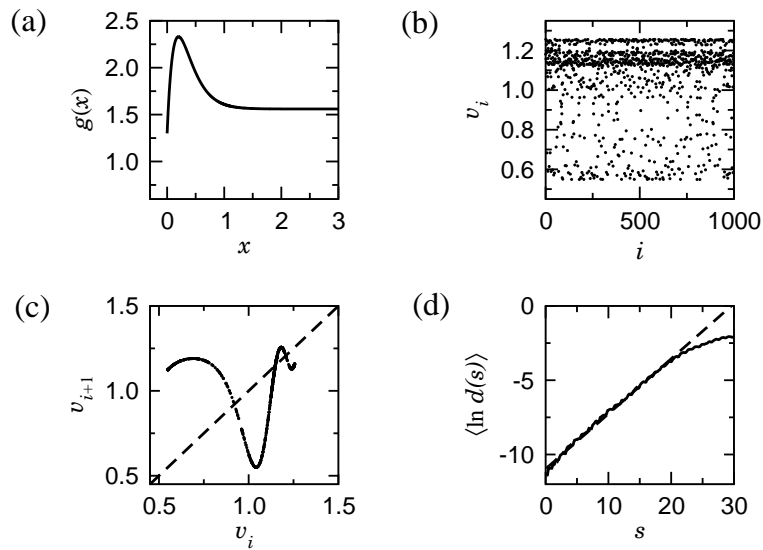}}
\end{center}
\caption{
	 (a) The degradation function $g(x)$ for the chaotic copier.
         (b) The sequence of values $v_i$ is chaotic.         
	 (c) The plot of $v_{i+1}$ {\it vs.} $v_i$ shows a clear 1D relation.
	 (d) Average of $\ln d(s)$ over 121 trajectories. The dashed line is a fit to the region of exponential scaling.
	 }
\label{fig:g_of_x}
\end{figure}
Figure~\ref{fig:g_of_x} shows the form of $g(x)$, the sequence of values $v_i$ that it generates through Eq.~(\ref{eqn:vmap}), and 
scatter plot of $v_{i+1}$ as a function of $v_i$. The latter falls on a 1D map given by  Eq.~(\ref{eqn:vmap}).
Note that neither $v$ nor $w$ ever takes on a value smaller than $\approx 0.5$, 
so the dynamics is consistent with a short-pulse
rejection mechanism with $\tspr$ less than this value. 
We have run the map for $10^6$ iterates and have not seen any evidence of periodicity. 
The Lyapunov exponent for 1D maps can be calculated by 
\begin{equation}
\lambda_{\mathrm{discrete}} = 
  \lim_{N\rightarrow \infty}\frac{1}{N}\sum^N_{i=1} \ln \left| \dfrac{dh(v_i)}{dv_i}\right|,
\label{eq:lyapunov}
\end{equation}
where $v_i$ is a fiducial trajectory, calculated numerically from the map evolution equation (Eq. (\ref{eqn:vmap})). 

With $N=10^4$ we obtained $\lambda_{\mathrm{discrete}} \approx 0.787$ for the same parameters used in Fig.~\ref{fig:g_of_x}. As $\lambda_{\mathrm{discrete}}$ gives the average expansion rate per cycle of the continuous system, we relate this Lyapunov exponent in discrete time to the maximum positive Lyapunov exponent $\lambda$ in continuous time by $\lambda = \lambda_{\mathrm{discrete}}/\overline{T}$, where $\overline{T}$ is the average cycle duration ($T_i = w_i+v_i$). 
Using this relation we can test the assumption of Zhang {\it et al.}\ (2009) that the Boolean distance $d(s)$ grows as $\exp(\lambda s)$. 
We measure $\overline{T} \approx 2.12$ from the discrete time series and, from the Boolean variable in continuous time, we calculate the average of $\ln d(s)$, shown in Figure \ref{fig:g_of_x}(d). In this section, however, instead of applying a perturbation to pairs of trajectories, as described in the previous section, we look for segments of the time series with an initially close Boolean difference, to reproduce the method in Zhang {\it et al.}\ (2009). In the region of exponential growth, $\left<\ln d(s) \right>$ is fit to a straight line with slope $\lambda = 0.37$, in good agreement with the ratio $\lambda_{\mathrm{discrete}}/\overline{T} \approx 0.371$. 
Details of the use of $d(s)$ for precise measurement of the largest Lyapunov exponent have yet to be worked out, but the good agreement between $\lambda$ and $\lambda_{\mathrm{discrete}}/\overline{T}$ indicates that the exponential growth of $d(s)$ is a reliable indicator of chaos.

Our simulations show that a simple copier with an appropriate form for the degradation effect can generate a chaotic sequence of pulse widths. We have observed chaos with other parameter values and other choices of the degradation function $g(x)$, but the analysis of the map in Eq.~(\ref{eqn:vmap}) for arbitrary $g(x)$ is difficult.  Two criteria are necessary for chaos (though not sufficient): the trajectory must not visit the vicinity of a stable fixed point of the map; and it must not visit $v\leq \tspr$ or a value of $v$ such that the corresponding $w\leq\tspr$ for whatever choice of $\tspr$ one takes to be of interest. Analytic expressions for these conditions are not easily determined.

In this section, we have shown that a simple, single-loop Boolean network (a copier with self-feedback) shows chaos when the degradation effect is taken into account.  In related work, chaos has also been observed in hybrid models with continuous variables governed by equations of the form $dx_i(t)/dt = F_i({\bf x}) - x_i(t)$, where $F$ is a binary-valued function (Mestl {\it et al.}\ 1996). From the perspective adopted in the present paper, such models may be thought of as explicit definitions of the underlying dynamics that produces degradation effects.  They are special cases, however, in that they do not include explicit time delays (or, alternatively, explicit descriptions of the propagation of signals along links).

Glass {\it et al.}\ (2005) have studied an electronic circuit explicitly designed to implement the hybrid model dynamics proposed as a model of a 5-node gene regulation network. In their circuit, the output of each logic gate is used to charge a capacitor, thereby simulating one specific form of a degradation effect.  They derive an approximate analytical solution for the response of the analog voltages in the circuit and show that the resulting map produces chaos.  
The 10 ms time constant introduced by the capacitor in their circuit is much slower than the switching times associated with the gates that implement the functions $F_i$ or the signal propagation times (typically on the order of tens of nanoseconds) and thus avoids the necessity of accounting for time delays explicitly.  Formally, this is equivalent to setting the time delay parameter $\tau$ in our model to zero.  The full dynamics of the hybrid system can be analyzed in this case without the need for a Boolean delay model of the type discussed in the present work.

Here, we treat the system variables as Boolean and account for the analog behavior through the effect of degradation on the time delays between gates.  Our method is potentially more flexible because we can incorporate a wider range of degradation functions that might arise in any particular model of the underlying dynamics and is applicable in cases where propagation delays are important.  Except for pulses near the short-pulse rejection limit, the variations due to degradation are small compared to the explicit delay $\tau$. Further investigation of the relation between the  circuit and models of (Glass {\it et al.}\ 2005) and (Mestl {\it et al.}\ 1996) is beyond our present scope, but may prove interesting.

\section{Boolean chaos in two loops}\label{sec:degradation_xor}
Returning to the \XOR\ system, we now show that a simple form of the
degradation effect can generate chaos there as well.  The situation is
not entirely analogous to the single copier studied in the previous
section because the two loops tend to create more pulses of rather short duration and
more possibilities for short-pulse rejections, both of which could
strongly influence the trajectory divergence rate.

\begin{figure}[htbp]
\begin{center}
\resizebox{\textwidth}{!}{\includegraphics{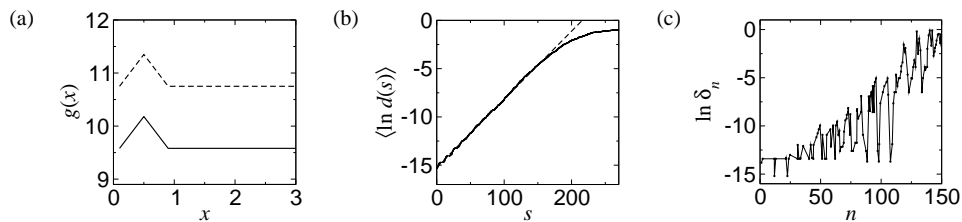}}
\end{center}
\caption{(a) Piecewise-linear degradation functions used on the two-loop system simulations. The solid and dashed line are the delays of links 1 and 2, respectively. (b) Average over 100 pairs of trajectories of the logarithm of the Boolean distance $d(s)$. The dashed line is a fit to a linear function of slope $\lambda = 0.071$.
(c) Logarithm of the timing difference $\delta_n$ between one pair of perturbed and unperturbed trajectories.  \label{fig:xordivergence}}
\end{figure}

In our simulations of the symmetric ABN with a single \XOR\, we use the degradation functions shown in Fig.~\ref{fig:xordivergence}(a) for the two links. Here, for ease of simulation, the delay times are given by a piecewise-linear function of the input pulse width. 
After calculating the time of a transition event using Eq.\ (\ref{eq:xor}) we calculate the input pulse width and correct the time delays according to the equation
\begin{equation}
g(x) = \left\{
	\begin{array}{lll}
	\tau_k +A (x-x_A), & \mathrm{for} & x_A < x \leq 0.50\\
	\tau_k -A (x-x_B), & \mathrm{for} & 0.50 < x \leq x_B\\
	\tau_k, & \mathrm{for} & x > x_B  
	\end{array}\right. ,
\label{eq:degradation}
\end{equation} 
where $x$ is a pulse width, $A= 1.50$, $x_A= 0.10$, $x_B= 0.90$, and $\tau_k$ assumes either the value $\tau_1 = 9.58$ or $\tau_2 = 10.75$, depending on which link is being traversed. Pulses shorter than $\tspr = 0.10$ are cutoff, as described previously.
We find that the details of the shape of the functions do not matter for the results presented below. 

Our simulations show that the system oscillates indefinitely without falling onto a periodic orbit or collapsing to the always-\OFF\ or -\ON\ states.  We use the two methods described in Sec.\ \ref{sec:divergence} to analyze the trajectory divergence. An extremely small perturbation $\epsilon = 10^{-6}$ was applied to a given orbit. Figure~\ref{fig:xordivergence}(b) shows the Boolean distance $d(s)$ averaged over 100 pairs, clearly indicating a substantial exponentially-increasing regime. 
Figure~\ref{fig:xordivergence}(c) shows the evolution of the difference $\delta_n=t'_n-t_n$ between one pair of original and disturbed trajectories. 
The difference $\delta_n$ increases exponentially, as  expected for an adequate definition of distance between trajectories in a chaotic system.
Therefore, we conclude that chaos, as defined by an exponential sensitivity to differences in the initial conditions, can be achieved by either one-loop or two-loop ABN only in the presence of degradation effects. Both $\delta_n$ and $d(s)$ are useful to distinguish the qualitative dynamics and to estimate the largest Lyapunov exponent of the system. Both quantities grow exponentially in the chaotic case and polynomially in the periodic case. 

\section{Conclusions}

Motivated by the recent experiments of Zhang \textit{et al.} (2009) where chaos was observed in an autonomous Boolean network, we studied systematically the dynamics of various simple Boolean networks.  From previous work on Boolean Delay Equations, it is known that Boolean networks whose nodes obey ideal Boolean rule display steady or periodic behavior, or display an ultraviolet catastrophe where the number of kinks circulating in the network per unit time grows as a power law. Hence, chaos is not possible for such a network. The ultraviolet catastrophe is obviously prevented in experiments due to non-ideal behavior of the logic gates.  These effects include short-pulse rejection, asymmetry between the logic states, and the degradation effect.  We first considered only the effect of short-pulse rejection for a network consisting of a single node and a single loop (link).  We proved that only periodic behavior is possible.  We then considered the case of a network consisting of a single node executing the \XOR\ function with two loops, which is known to display an ultraviolet catastrophe. Even in this case, short-pulse rejection renders the behavior periodic.  The situation is less clear when we take into account both short-pulse rejection and asymmetry between the logic states.  While we were unable to prove that the two-loop network is always periodic, numerical simulations suggest that this is the case. Finally, we showed, through numerical simulations, that chaos is possible for even the simplest network consisting of a copier and a single loop when the degradation effect is included in the model.  Chaos is also displayed in the two-loop \XOR\ network when the degradation effect is taken into account.  
Given that a degradation effect is present at some level in any real network, our results strongly suggest that there exists a class of experimental Boolean-like networks, containing at least one \XOR\ connective and feedback loop, whose elements would display deterministic chaos.

\section{Acknowledgements}
We thank Andrew Mullhaupt for discussions of this work.  JESS gratefully acknowledges the financial support of the NSF under grant PHY-0417372.  HLDSC, DJG and RZ gratefully acknowledge the financial support of the US Office of Naval Research under MURI award \#N00014-07-1-0734.

\end{document}